%% file: mydocument.tex
\pdfoutput=1
\newcommand*{\ATLASLATEXPATH}{}
\documentclass[atlasdraft=false, texlive=2016, UKenglish, LANGEDIT=false,cernpreprint]{\ATLASLATEXPATH atlasdoc}
\usepackage{\ATLASLATEXPATH atlaspackage}
\usepackage{booktabs}
\usepackage{\ATLASLATEXPATH atlasbiblatex}
\usepackage{\ATLASLATEXPATH atlasphysics}
\usepackage{tikz}
\usetikzlibrary{decorations.markings}
\usetikzlibrary{snakes}
\tikzset{
    fermion/.style={draw=black, postaction={decorate},
        decoration={markings,mark=at position .55 with {\arrow[draw=black]{>}}}},
     gluon/.style={decorate, draw=black,
        decoration={coil,amplitude=3pt, segment length=5pt}},
}
\usepackage{atlas-paper-defs}

\AtlasTitle{Properties of $g\rightarrow b\bar{b}$ at small opening angles in $pp$ collisions with the ATLAS detector at $\sqrt{s}=13$~\TeV}

\AtlasAbstract{
The fragmentation of high-energy gluons at small opening angles is largely unconstrained by present measurements.  
Gluon splitting to $b$-quark pairs is a unique probe into the properties of gluon fragmentation because identified $b$-tagged jets 
provide a proxy for the quark daughters of the initial gluon.  In this study, key differential distributions related to 
the $g\rightarrow b\bar{b}$ process are measured using $33$ fb$^{-1}$ of $\sqrt{s}=13$~\TeV\ $pp$ collision data 
recorded by the ATLAS experiment at the LHC in 2016.  Jets constructed from charged-particle tracks, clustered 
with the anti-$k_t$ jet algorithm with radius parameter $R = 0.2$, are used to probe angular scales below the $R=0.4$ jet radius.  
The observables are unfolded to particle level in order to facilitate direct comparisons with predictions from present and future simulations.  
Multiple significant differences are observed between the data and parton shower Monte Carlo predictions, providing 
input to improve these predictions of the main source of background events in analyses involving boosted Higgs bosons decaying into $b$-quarks.
}

\PreprintIdNumber{CERN-EP-2018-323}
\AtlasJournal{Phys. Rev. D.}

\hypersetup{pdftitle={ATLAS document},pdfauthor={The ATLAS Collaboration}}

\begin{document}
\maketitle

\tableofcontents

\section{Introduction}
\label{sec:intro}

The search for highly Lorentz-boosted Higgs bosons produced by Standard Model (SM) processes~\cite{Sirunyan:2017dgc} or by beyond-the-SM (BSM) processes~\cite{Aaboud:2017ecz,Aaboud:2017ahz,Aaboud:2017yqz,Aaboud:2016xco,Khachatryan:2016cfa,Sirunyan:2017nrt} is of crucial importance at the Large Hadron Collider (LHC).  As the branching ratio for the Higgs boson to decay into bottom quark pairs dominates the total decay rate, the boosted $H\rightarrow b\bar{b}$ channel can be the most sensitive to BSM effects entering at high Higgs boost.  Algorithms for identifying jets resulting from bottom quark fragmentation are very powerful, so the main background for searches with boosted Higgs bosons contains $b$-quarks.  The main contribution to this background is gluon splitting to $b\bar{b}$ pairs at small opening angles since the angle between the $b$-quarks in $H\rightarrow b\bar{b}$ scales with the Higgs boson mass ($m_H$) and momentum ($p_H$) as $m_H/p_H$.  The $g\rightarrow b\bar{b}$ process also contributes to many other important SM measurements and searches by providing a source of additional real $b$-quark jets that can fake a signal for $b$-quarks originating from other processes (see e.g.\ Refs.~\cite{SUSY-2016-10,Aaboud:2018bir,Aaboud:2017rss,ATLAS-CONF-2018-0362,CMS-PAS-HIG-18-016}).  

The modeling of $g\rightarrow b\bar{b}$ fragmentation is complex and provides a useful probe of quantum chromodynamics (QCD).  The large mass of the $b$-quark introduces a significant modification to the massless QCD splitting functions by screening the soft-emission singularity.  Trijet measurements from LEP~\cite{Barate:1998vs,Abbiendi:2000zt,Abreu:1999qh} and SLD~\cite{Abe:2001csa} provide valuable information about the rate of $g\rightarrow b\bar{b}$, but have not explored the differential properties of the fragmentation in the small opening-angle regime.  Previous measurements that include the $b\bar{b}$ final state at the S$p\bar{p}$S, Tevatron, and LHC using inclusive~\cite{Albajar:1993be,Albajar:1990zu,Aaboud:2017vqt,Khachatryan:2011wq,Chatrchyan:2012hw,Aad:2011rr,Abbott:1999se,Aaltonen:2007zza,Abbott:1999wu,Abachi:1994kj,Khachatryan:2011wq,Aaij:2017pvu,Aaij:2010gn,Aaij:2013noa,Aaij:2012jd,Aaij:2010gn,Aad:2012jga,Aad:2011sp,Aad:2015duc,Khachatryan:2010yr,Acosta:2004yw,ATLAS:2013cia}, multijet~\cite{Aaboud:2016jed,ATLAS:2011ac,Chatrchyan:2012dk}, and associated production~\cite{Aad:2013vka,Aad:2014dvb,Chatrchyan:2014dha,Abazov:2014fka,Abazov:2013uza,Abazov:2015aha,Aaltonen:2013ama,Aaltonen:2009qi,Aaltonen:2008mt,Chatrchyan:2013zja,Chatrchyan:2013zja} topologies have focused on well-separated quark pairs (dominated by fixed-order instead of parton-splitting effects) and were limited in their kinematic reach due in part to small datasets and low momentum transfers.

\begin{figure}[h!]
\centering
\begin{tikzpicture}[line width=1.5 pt,scale=0.8]
	\draw[fermion] (-1,-2.5-0.25+1.5) -- (1,-0.5-0.25+1.5);
	\draw[gluon] (-1,2.5+0.25) -- (1,0.5+0.25);
	\draw[fermion] (1,0.75) -- (3,0.75);
	\draw[fermion] (1+2,-0.5-0.25+1.5) -- (3+2,-2.5-0.25+1.5);
	\draw[gluon] (1+2,0.5+0.25) -- (2+2,1.5+0.25);
	\draw[fermion] (2+2,1.5+0.25)-- (3+2,2.5+0.25);
	\draw[fermion] (3+2,0.5+0.25)-- (2+2,1.5+0.25);
	\node at (-1.3, -2.8+1.5) {\huge $q$};
	\node at (-1.3, 2.8) {\huge $g$};
	\node at (3.3+2, -2.8+1.5) {\huge $q$};
	\node at (3.3+2, 2.8) {\huge $b$};	
	\node at (3.3+2, 0.8) {\huge $\bar{b}$};	
 \end{tikzpicture}
 \caption{A representative diagram for the high-$p_\text{T}$ $g\rightarrow b\bar{b}$ process studied in this paper.}
  \label{fig:feynman}
 \end{figure}
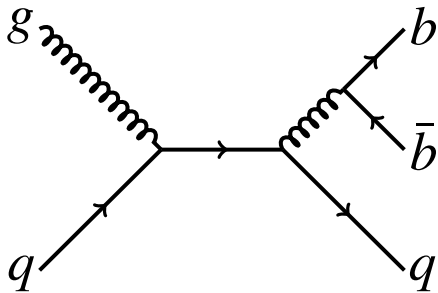

The high transverse momentum and low angular separation regime for $g\rightarrow b\bar{b}$ can be probed at the LHC using $b$-tagged small-radius jets within large-radius jets. This topology is used to calibrate $b$-tagging in dense environments~\cite{ATLAS-CONF-2016-039,ATLAS-CONF-2016-002,CMS:2013vea} and is studied phenomenologically~\cite{Anderle:2017qwx,Ilten:2017rbd}.  
The measurement shown in this paper builds on these studies by using data collected by the ATLAS detector from $\sqrt{s}=13$~\TeV\ $pp$ collisions in order to perform a differential cross-section measurement of $g\rightarrow b\bar{b}$ inside jets at high transverse momentum -- see Figure~\ref{fig:feynman} for a representative Feynman diagram. Small-radius jets built from charged-particle tracks are used as proxies for $b$-quarks and can be used as precision probes of the small opening-angle regime. 

This paper is organized as follows.  After a brief introduction to the ATLAS detector in Section~\ref{sec:detector}, the data and simulations used for the measurement are documented in Section~\ref{sec:data}.  Section~\ref{sec:obj} describes the event selection and Section~\ref{sec:obs} lists and motivates the observables to be measured.  The key challenge in the measurement is the estimation of background processes, which is performed using a data-driven approach illustrated in Section~\ref{sec:analysis}.  The data are unfolded to correct for detector effects to allow
direct comparisons to particle-level predictions.  This procedure is explained in Section~\ref{sec:unfold} and the associated systematic uncertainties are detailed in Section~\ref{sec:systs}.   The results are presented in Section~\ref{sec:result} and the paper concludes with Section~\ref{sec:conclusion}.

\section{ATLAS detector}
\label{sec:detector}

The ATLAS detector~\cite{PERF-2007-01} is a multipurpose particle detector with a forward/backward-symmetric cylindrical geometry. The detector has a nearly $4\pi$ coverage in solid angle\footnote{ATLAS uses a right-handed coordinate system with its origin at the nominal interaction point (IP) in the center of the detector and the $z$-axis along the beam pipe. The $x$-axis points from the IP to the center of the LHC ring, and the $y$-axis points upward. Cylindrical coordinates ($r$, $\phi$) are used in the transverse plane, $\phi$ being the azimuthal angle around the beam pipe. The pseudorapidity is defined in terms of the polar angle as $\eta = -\ln \tan(\text{polar angle}/2)$.} and consists of an inner tracking detector, electromagnetic and hadronic calorimeters, and a muon spectrometer. The inner detector (ID) is surrounded by a superconducting solenoid providing a \SI{2}{T} magnetic field and covers a pseudorapidity range of $|\eta| < 2.5$. The ID is composed of silicon pixel and microstrip detectors as well as a transition radiation tracker. For the LHC $\sqrt{s}=\SI{13}{\TeV}$ run, the silicon pixel detector has been upgraded to include an additional layer close to the beam interaction point~\cite{Abbott:2018ikt}. The lead/liquid-argon electromagnetic sampling calorimeters measure  electromagnetic energies 
with high granularity for the pseudorapidity region of $|\eta| < 3.2$. Hadron energies are measured by a hadronic (steel/scintillator tile) calorimeter with $|\eta| < 1.7$. 
The forward and endcap regions between $1.5<|\eta| <4.9$ are instrumented with liquid-argon calorimeters for both the electromagnetic and hadronic measurements. Surrounding the calorimeters, the muon spectrometer includes three large superconducting toroidal magnets with eight coils each. The muon spectrometer has a system of precision tracking chambers covering $|\eta| < 2.7$ and fast trigger chambers covering $|\eta| < 2.4$. A two-level trigger system is used to select events for further analysis~\cite{TRIG-2016-01}. The first-level trigger is implemented in hardware and utilizes partial detector information to reduce the accepted event rate to \SI{100}{kHz}. The high-level trigger is based on software and accepts events at a rate of \SI{1}{kHz}.

\section{Datasets}
\label{sec:data}

This measurement uses the dataset of $pp$ collisions recorded by the ATLAS detector in 2016, corresponding to an integrated luminosity of 33~\ifb\ at a center-of-mass-energy of $\sqrt{s}=13$~\TeV.   Events are considered only if they were collected during stable beam conditions and satisfy data quality requirements. Due to the high instantaneous luminosity and the large total inelastic proton--proton scattering cross section, on average there are about 25 simultaneous (pileup) collisions in each bunch crossing.

The measurement presented in this paper uses  a variety of Monte Carlo (MC) simulated event samples to correct for detector effects as well as for direct comparisons with the unfolded data. Inclusive jet events were generated at leading order in the strong coupling constant, $\alphas$, with \PYTHIAV{8.1}~\cite{Sjostrand:2007gs} using a $2 \rightarrow 2$ matrix element, the NNPDF2.3LO PDF set~\cite{Ball:2012cx}, and a set of generator parameter values called the A14 tune~\cite{ATL-PHYS-PUB-2014-021}.  The \textsc{EvtGen}~\cite{Lange:2001uf} program was used to model the heavy-flavor decays to agree  with experimental data.  Additional inclusive jet events were simulated using a different generator in order to study the impact of modeling uncertainties related to both the perturbative and non-perturbative aspects of fragmentation. \SHERPAV{2.1}~\cite{Gleisberg:2008ta} generates events using multi-leg $2\rightarrow N$ matrix elements, which are matched to parton showers following the CKKW prescription~\cite{Catani:2001cc}. These \SHERPA events were simulated using the CT10 PDF set~\cite{Lai:2010vv} and the default parameter tune in \SHERPA.  

Energy depositions from particles in MC event samples interacting with the detector and the subsequent detector readout were modeled using a full simulation of the ATLAS detector~\cite{SOFT-2010-01} implemented in Geant4~\cite{Agostinelli2003250}.  The effects of pileup were simulated with unbiased $pp$ collisions using \PYTHIAV{8.1} and overlaid on the nominal inclusive jet events.

\section{Object and event selection}
\label{sec:obj}

Section~\ref{sec:objectdef} describes the definition of collision vertices, charged-particle tracks, jets, and $b$-jets.  Following the discussion of objects, Section~\ref{sec:events} describes the particle-level definition of the measurement phase space and the detector-level selections used to create an enriched sample of jets resulting from $g\rightarrow b\bar{b}$.

\subsection{Object reconstruction, identification and association}
\label{sec:objectdef}

Collision vertices are reconstructed from ID tracks. Each vertex is required to be associated with at least two tracks with $p_\text{T}>0.4$~\GeV. The primary hard-scattering vertex of the event is chosen to be the vertex having the highest $\sum p_\text{T}^{2}$ calculated using all associated tracks.  Particle-level events in simulation are considered before the addition of pileup and therefore there is no ambiguity in selecting the collision vertex.

Calorimeter jets are built from calibrated topological calorimeter-cell clusters~\cite{PERF-2014-07} using the anti-$k_t$~\cite{Cacciari:2008gp} algorithm with radius parameter $R=1.0$ as implemented in FastJet~\cite{Cacciari:2011ma}. Jets are groomed using a trimming procedure~\cite{Krohn:2009th}.  This procedure reclusters the constituents of a jet into subjets with a smaller radius of size $R_\text{sub}=0.2$ and removes those subjets with a low fraction of the full jet momentum $f_\text{cut}=0.05$.  Following jet grooming, the mass and momentum of the resulting jets are corrected so that the detector-level values match the particle-level values on average~\cite{ATLAS-CONF-2016-035}.  These large-radius jets are proxies for the gluons.  

Smaller-radius jets are used as proxies for the $b$-quarks originating from the gluons.  For this purpose, jets are clustered using tracks as inputs (\textit{track-jets}).  Track-jets clustered from tracks with $p_\text{T}>500$~\MeV\ that have been well-matched to the primary vertex\footnote{The transverse impact parameter must be less than 2~mm and the longitudinal impact parameter multiplied by the sine of the polar angle must be less than 3~mm.} and are composed of at least one pixel detector hit and at least six hits in the silicon tracker~\cite{ATL-PHYS-PUB-2015-051} are constructed using the anti-$k_t$ algorithm with $R=0.2$.  Track-jets are required to have at least two tracks.  Four-vectors are calculated for each track, assuming the mass of the charged pion.
Small-radius jets are matched to large-radius jets via ghost association~\cite{Cacciari:2008gn}.  This matching procedure creates ghost versions of the small-radius jets with the same direction but infinitesimal $p_\text{T}$.  Jet clustering is repeated and small-radius jets are assigned to the large-radius jet that contains their ghosted version.  Since the jet finding algorithm is infrared safe, the four-momenta of the jets are unaffected by the addition of ghosts.

Particle-level jets are clustered using the same algorithms as for detector-level jets, except the inputs to jet finding are all stable particles ($c\tau > 10$~mm) excluding all muons and neutrinos.  The same trimming algorithm applied to calorimeter jets is also applied to the large-radius particle-level jets.  Particle-level track-jets are formed from all stable charged particles that have $p_\text{T}>500$~\MeV\ and $|\eta|<2.5$, excluding muons.

Track-jets that are likely to have originated from the fragmentation of a $b$-quark ($b$-jet) are identified using the \textit{MV2c10} algorithm~\cite{PERF-2012-04,ATL-PHYS-PUB-2016-012}, which is a combination of three baseline algorithms \textit{IP3D}, \textit{SV}, and \textit{JetFitter}. The \textit{IP3D} algorithm uses log-likelihood ratios of the three-dimensional signed impact parameter significance of tracks associated with jets. The \textit{SV} tagger reconstructs the secondary decay vertices of $b$-hadrons. The \textit{JetFitter} algorithm reconstructs the topology of detached vertices along the $b$-hadron decay axis. Finally, the $MV2c10$ algorithm combines the outputs of the baseline algorithms with a boosted decision tree and assigns a probability of a jet being a $b$-jet, $c$-jet, or light-flavor jet. The selected working point corresponds to a $b$-jet efficiency, a $c$-jet efficiency, and a light-flavor-jet rejection of $\epsilon_b = 60\%, \epsilon_c=15\%$, and $1/\epsilon_\text{light}=480$, respectively, as measured in $t\bar t$ events for jets with $p_\text{T}>10$~\GeV\ and $|\eta|<2.5$. A scale factor is applied to MC events to account for the measured efficiency difference between data and MC events at the chosen working point~\cite{Aaboud:2018xwy}.   

Particle-level track-jets are tagged as $b$-jets if there is a $b$-hadron from the simulated event record with $p_\text{T}>5$~\GeV\  that is ghost-associated with the jet.  If instead a hadron containing a $c$-quark from the event record with $p_\text{T}>5$~\GeV\ can be matched to the jet by ghost-association, it is declared a $c$-jet.  All other jets are declared to be light-flavor jets. 

\subsection{Event selection}
\label{sec:events}

At detector level events are selected using single-jet triggers.  In the first-level trigger, a  sliding-window algorithm based on low-granularity calorimeter towers records events with transverse energy greater than $100$~\GeV.  In the high-level trigger, $R=1.0$ anti-$k_t$ jets are formed from calibrated calorimeter-cell clusters and the $p_\text{T}$ threshold is 420~\GeV.  This trigger scheme is fully efficient for calibrated offline jets with $p_\text{T}>450$~\GeV\ and $|\eta| < 2$ and therefore these kinematic requirements are used to select jets for the measurement.  
The offline analysis requires the highest-$p_\text{T}$ calorimeter jet to have at least two
associated track-jets with $p_\text{T}>10$~\GeV\ and $|\eta|<2.5$.  If there are more than two track-jets, only the leading two are used for subsequent analysis.  In order to enhance the $g\rightarrow b\bar{b}$ purity, the leading track-jet associated with the selected calorimeter jet must be $b$-tagged by the \textit{MV2c10} algorithm at the $60\%$ efficiency working point.  Requiring both track-jets to be $b$-tagged increases the purity but degrades the precision of the background fit described in Section~\ref{sec:analysis} and so only one is required.

At particle level events are required to have at least one large-radius jet with $p_\text{T}>450$~\GeV. The leading jet needs to have at least two associated particle-level track-jets with $p_\text{T}>10$~\GeV.  Both of the associated small-radius jets must be tagged as $b$-jets.

This inclusive event selection produces a sample where QCD scattering processes dominate.

\section{Observables}
\label{sec:obs}

The kinematic properties of the $g\rightarrow b\bar{b}$ process are characterized by three quantities: the opening angle between the $b$-quarks, the momentum sharing between the $b$-quarks, and the orientation of the gluon splitting relative to the gluon production plane.  The first of these quantities is probed by measuring the $\Delta R(b,b)=\sqrt{\Delta\phi(b,b)^2+\Delta\eta(b,b)^2}$ between track-jets.  Momentum sharing is explored using the quantity $z(p_\text{T})=p_\text{T,2}/(p_\text{T,1}+p_\text{T,2})$, where $p_\text{T,1}$ and $p_\text{T,2}$ are the transverse momenta of the leading and subleading track-jets, respectively.  A quantity sensitive to the relative orientation of the gluon splitting is $\Delta\mathrm{\theta}_\text{ppg,gbb}$, which is the angle between the plane spanned by the beam-line and the vector sum of the two track-jets and the plane spanned by the two track-jets.  The angular quantities $\Delta R(b,b)$ and $\Delta\mathrm{\theta}_\text{ppg,gbb}$ are depicted in Figure~\ref{fig:schematic}.  

In addition to these quantities, the dimensionless mass $\log(m_{bb}/p_\text{T})$ is also measured, where the mass and $p_\text{T}$ in the logarithm are computed from the four-vector sum of the two track-jets.  The $b \bar b$ mass is an important observable for measurements and searches with Higgs and $Z$ bosons.   Track-jets, in contrast to the calorimeter-based subjets from trimming, are used due to their excellent angular resolution; in simulation, there is little difference between using the directions from the track-jets, the full jets (including neutrals), or the $b$-hadrons from the $b$-quark fragmentation.  There is some discrepancy between these different objects for the energy-dependent observables, but the track-jets are still useful due to their excellent momentum resolution in the gluon $p_\text{T}$ range probed in this measurement.   The excellent angular and momentum resolutions are presented in Figure~\ref{fig:response}, which shows the detector response for all four observables targeted with this measurement.

\begin{figure}[htbp]
  \centering
 \includegraphics[width=0.8\textwidth]{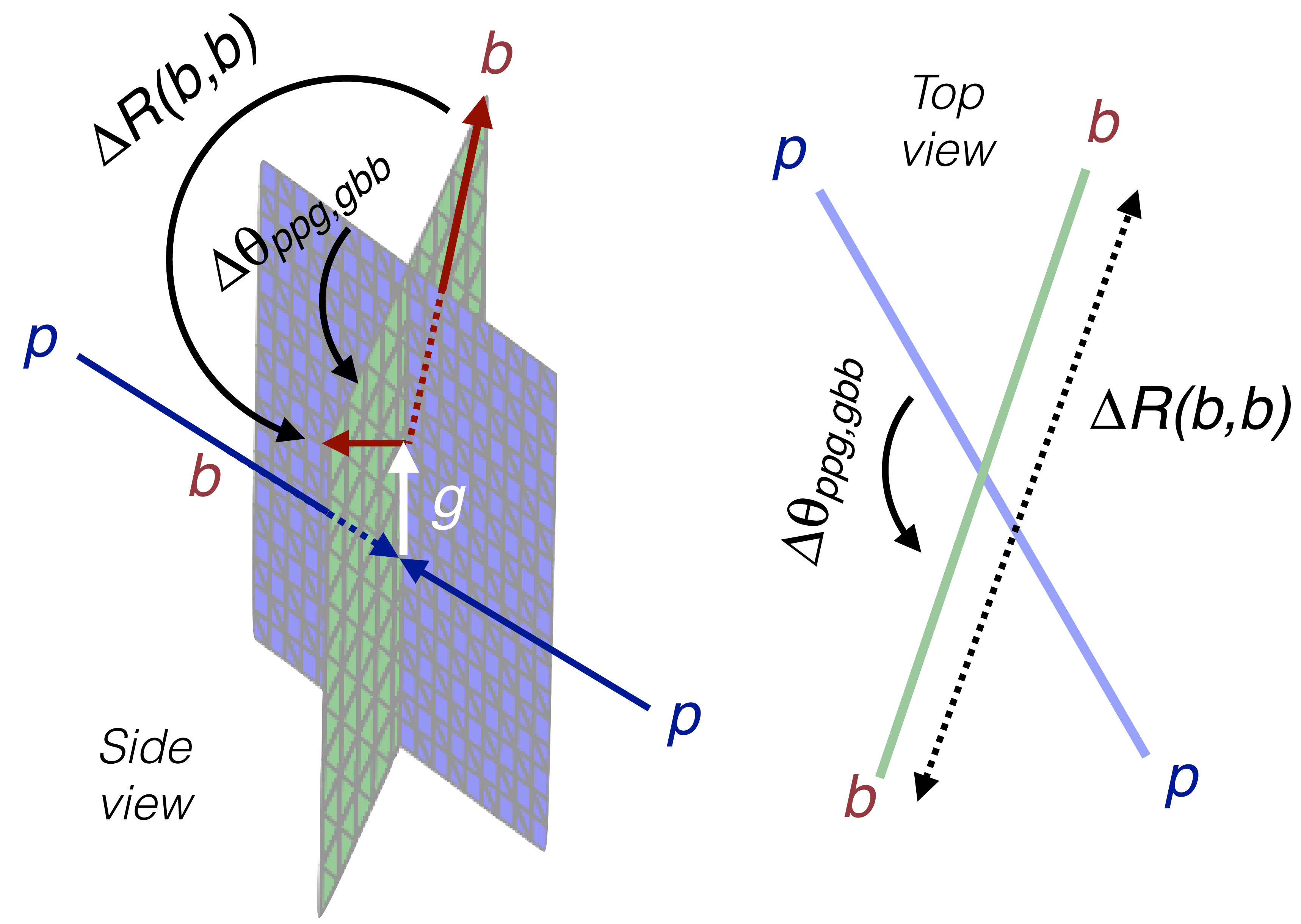}
\caption{Schematic diagrams illustrating the $\Delta R(b,b)$ and $\Delta\mathrm{\theta}_\text{ppg,gbb}$ observables. In
this example, the gluon is emitted at $\eta=0$.}
  \label{fig:schematic}
\end{figure}

\begin{figure}[htbp]
  \centering
 \includegraphics[width=0.48\textwidth]{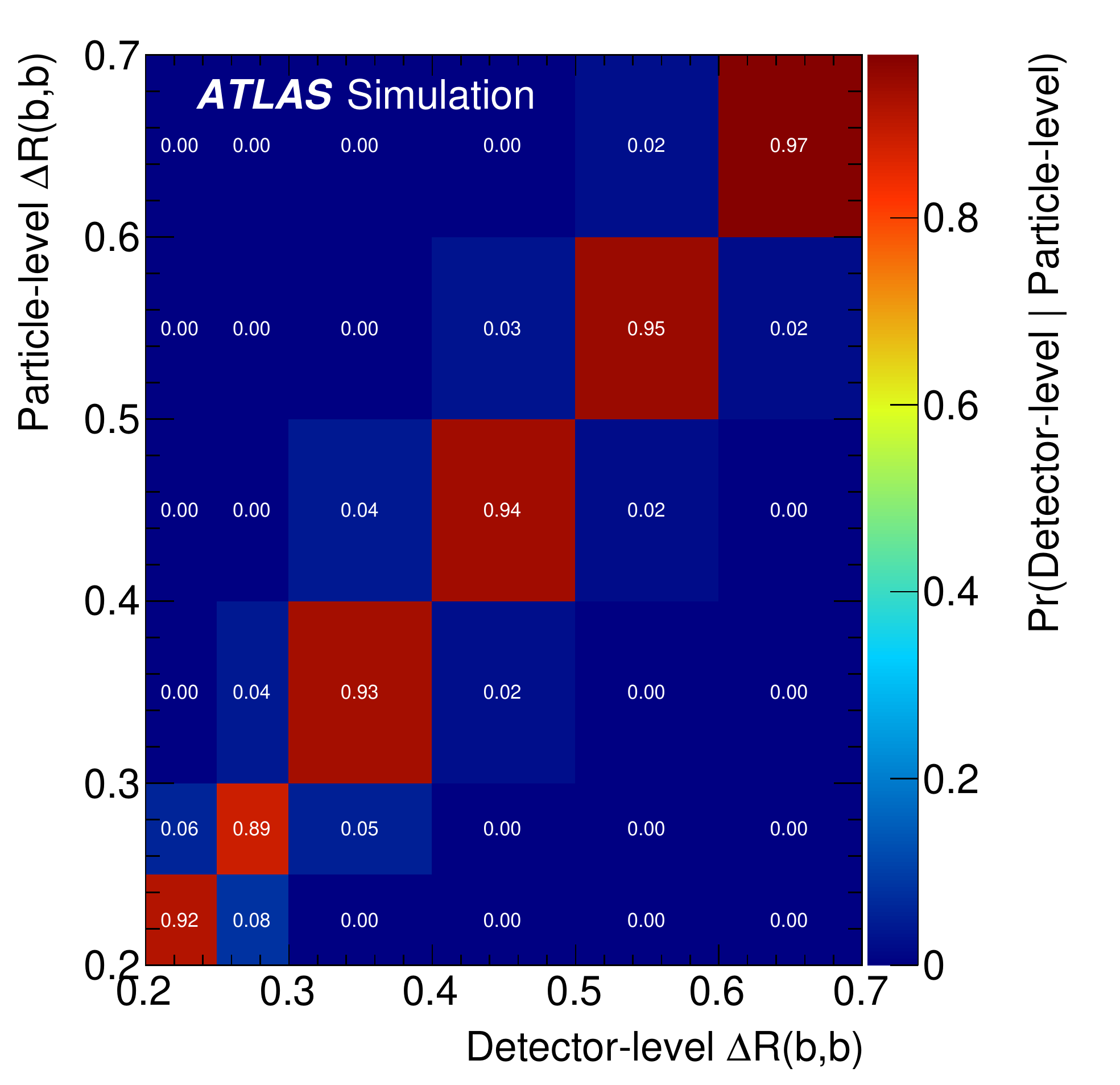} \includegraphics[width=0.48\textwidth]{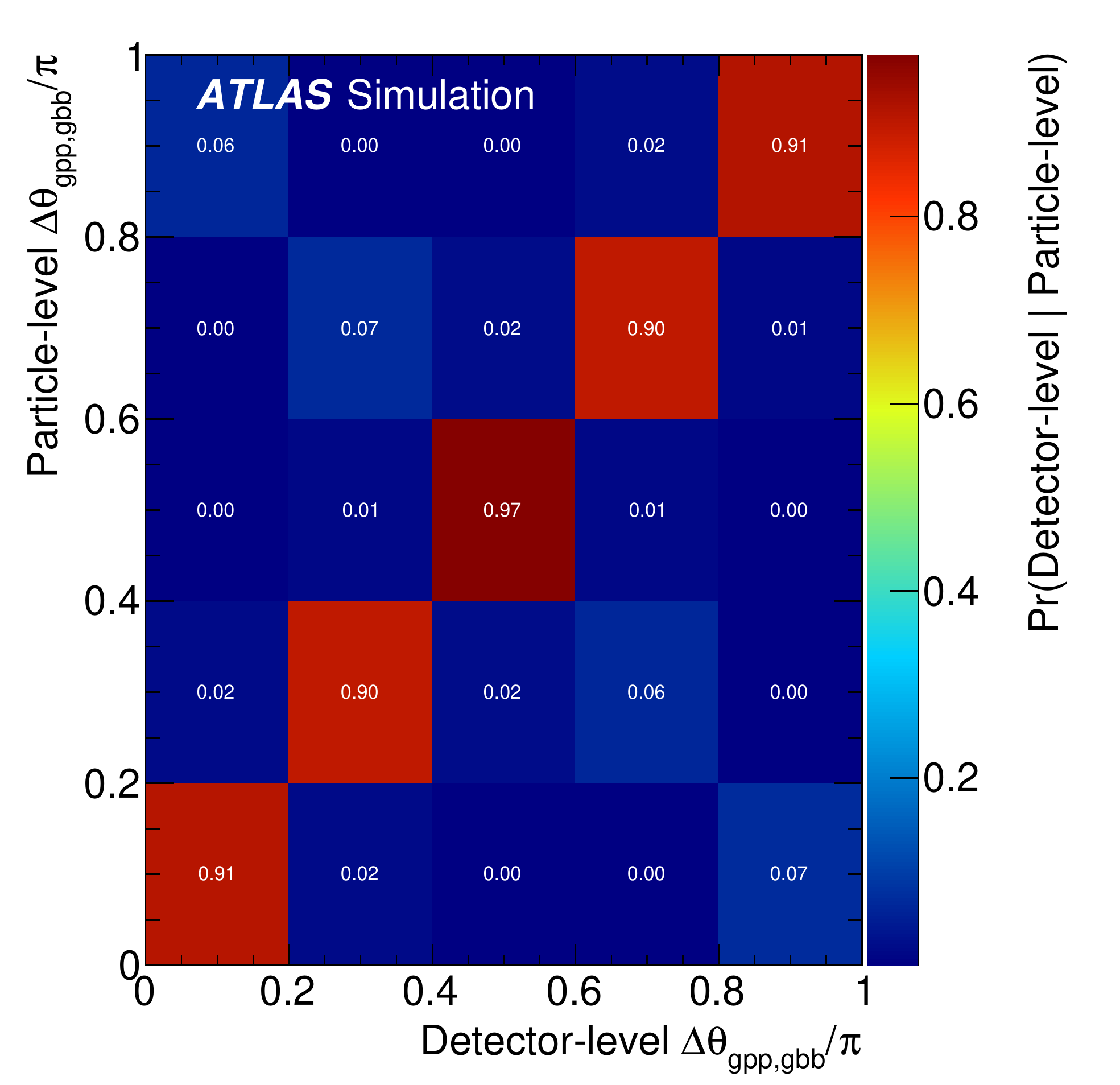}\\
  \includegraphics[width=0.48\textwidth]{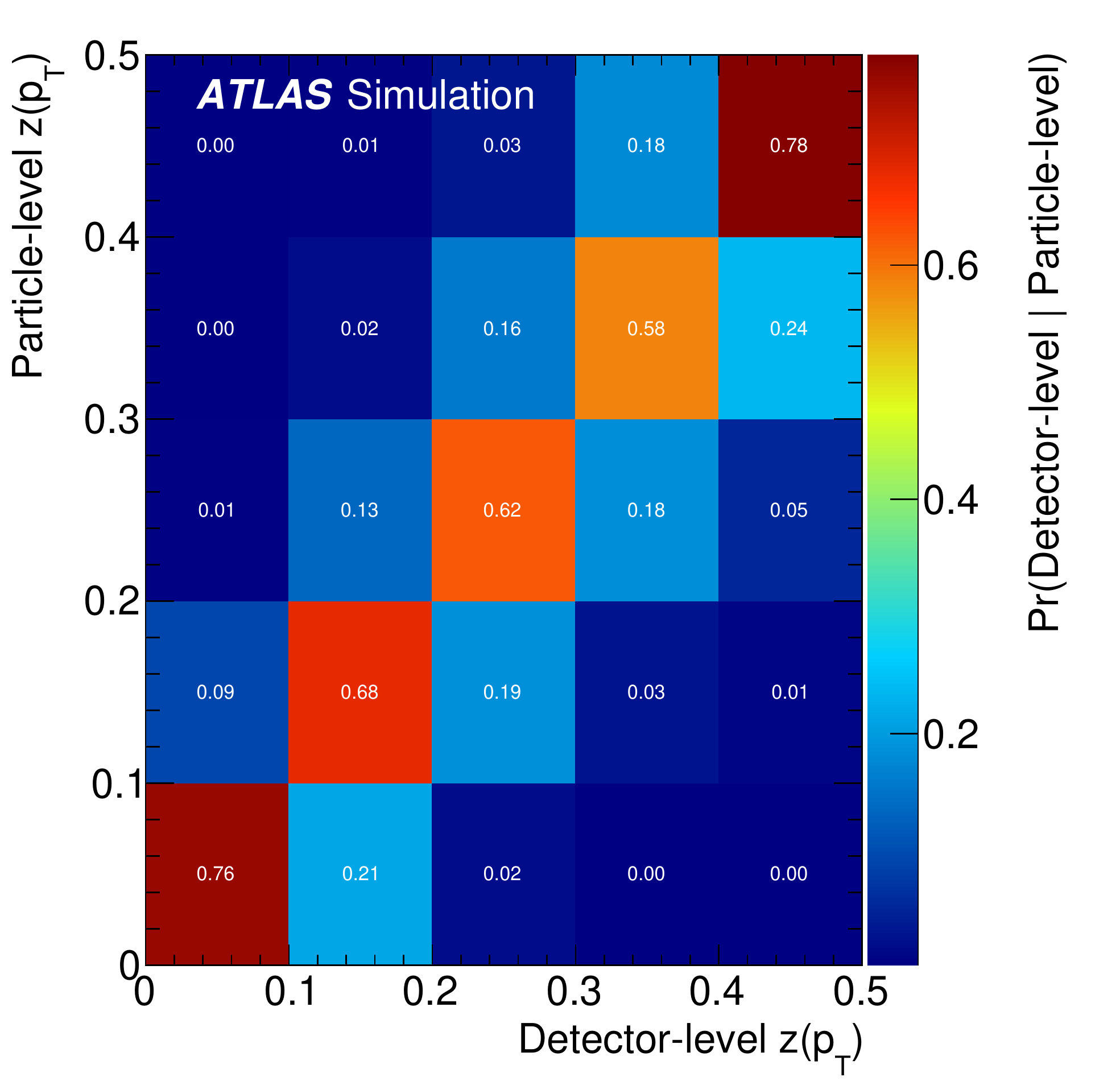} \includegraphics[width=0.48\textwidth]{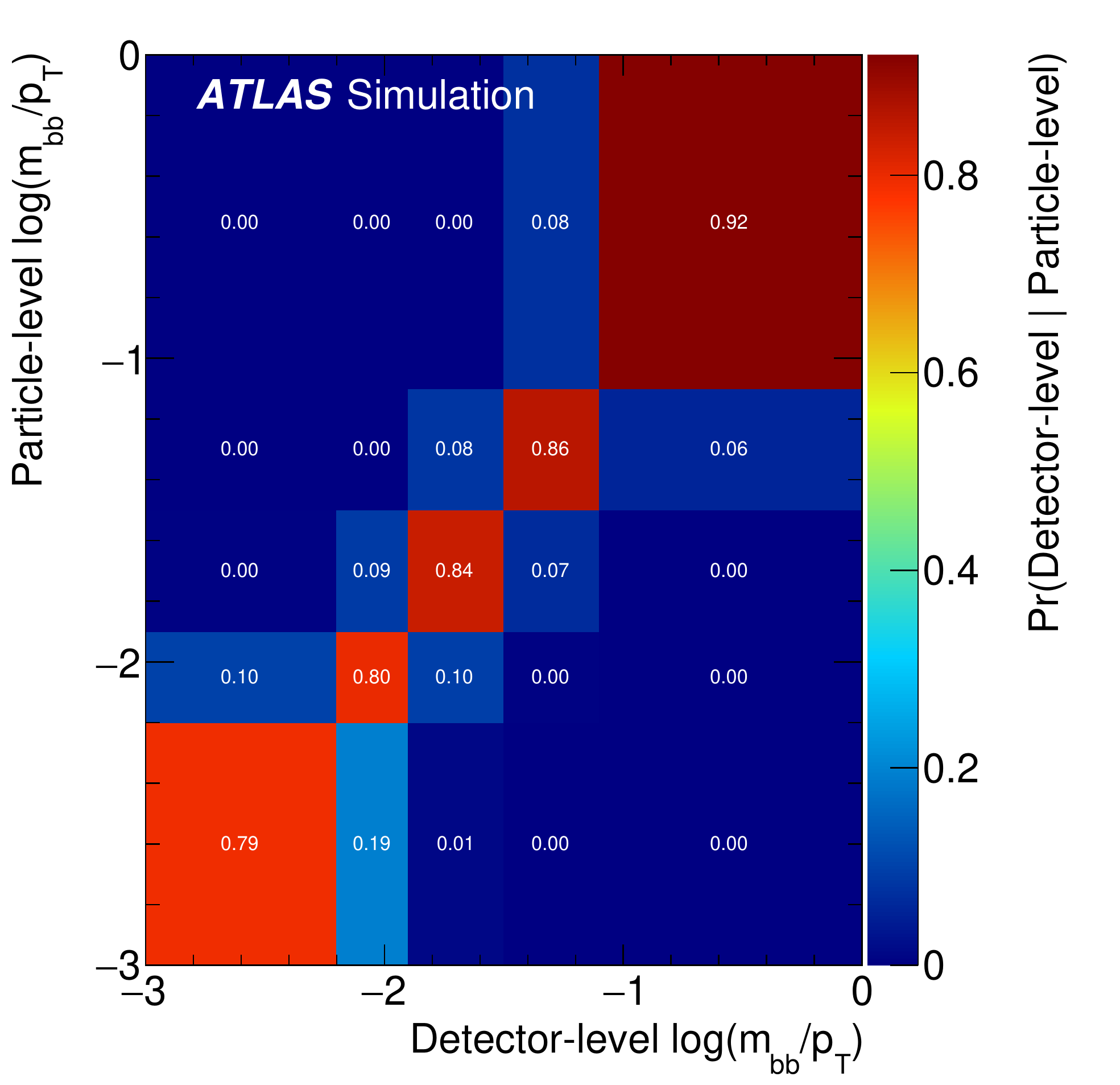}
\caption{The detector response is represented as the conditional probability of the detector-level quantity given the particle-level quantity, written as $\Pr(\text{detector-level}|\text{particle-level})$, in simulation for $\Delta R(b,b)$ (top left), $\Delta\mathrm{\theta}_\text{ppg,gbb}$ (top right), $z(p_\text{T})$ (bottom left), and $\log(m_{bb}/p_\text{T})$ (bottom right).  The small anti-diagonal component for $\Delta\mathrm{\theta}_\text{ppg,gbb}$ is due to cases where the leading and subleading track-jets are swapped between detector level and particle level so $\Delta\mathrm{\theta}_\text{ppg,gbb}\mapsto \pi-\Delta\mathrm{\theta}_\text{ppg,gbb}$. }
  \label{fig:response}
\end{figure}

\section{Background estimation}
\label{sec:analysis}

After the event selection, the contribution from large-radius jets that do not have two associated track-jets containing B-hadrons is subtracted from data, as described below, before correcting for detector effects.
The fraction of background events may not be well-modeled by the simulation, so correction factors are determined from data template fits to the impact parameter distribution and applied for each bin of the four target observables of the analysis
prior to subtraction.  In each bin of the target observable distributions, the distribution of the signed impact parameter significance $s_{d_0}$ is fitted to data using templates from simulation while letting the fraction of each flavor component float in the fit~\cite{ATLAS-CONF-2016-039,ATLAS-CONF-2016-002}.  For a given track, $s_{d_0}=s_j|d_0|/\sigma(d_0)$, where $d_0$ is the transverse impact parameter relative to the beam-line and $\sigma(d_0)$ is the uncertainty in $d_0$ from the track fit and the variable $s_j$ is the sign of $d_0$ with respect to the jet axis: $s_j=+1$ if $\sin(\phi_\text{jet}-\phi_\text{track})\cdot d_0>0$ and $s_j=-1$ otherwise.    The transverse impact parameter itself is signed, with $\text{sign}(d_0) = \text{sign}(\vec{p}_\text{T,track}\times(\vec{r}_\text{IP,xy}-\vec{r}_\text{PV,xy}))$, where $\vec{r}_\text{IP,xy}$ and $\vec{r}_\text{PV,xy}$ are the locations of the track impact parameter and primary vertex, respectively, in the transverse plane.

Due to the long lifetime of $b$-hadrons, the values of $s_{d_0}$ for tracks in $b$-jets tend to be larger than those for tracks in $c$-jets and light-flavor jets.  Therefore, the distribution of $s_{d_0}$ can be used to extract the fractions of $b$-jets, $c$-jets, and light-flavor jets using templates from simulation.  For each track-jet $j_i$, the $s_{d_0}$ from the track with the second largest $|s_{d_0}|$, called $\subsdzero(j_i)$, is used for the extraction.  The leading and third-leading $s_{d_0}$ values (ordered by $|s_{d_0}|$) are used as a validation of the flavor-fraction fitting procedure and produce consistent results.  The leading $s_{d_0}$ is not as well-modeled as $\subsdzero$ and therefore the $\chi^2$ resulting from the fit procedure described below is slightly worse.  The value of $\subsdzero$ does not have a strong dependence on jet $p_\text{T}$, so the fit is performed inclusively.  This choice was validated by using $p_\text{T}$-binned fits, which produce results consistent with the inclusive approach.

A binned maximum-likelihood fit to $\subsdzero(j_1)$ and $\subsdzero(j_2)$ is performed to extract the flavor fractions.  Given the flavors of track-jets $j_1$ and $j_2$ ($p_\text{T,1}\geq p_\text{T,2}$), $\subsdzero(j_1)$ and $\subsdzero(j_2)$ are well-approximated as being statistically independent (linear correlation is less than 5\%).  Therefore, the probability distribution $p(\subsdzero(j_1), \subsdzero(j_2))$ can be approximated by the product of marginals $p(\subsdzero(j_1))\times p(\subsdzero(j_2))$. This approximation reduces a two-dimensional fit to a simultaneous fit of two one-dimensional distributions.  In order to increase the robustness of the fit, flavor combinations with similar templates are merged.  The three templates used for each bin of the target observable are BB (signal), B, and L+C.   The BB template only includes events where both jets are labeled as $b$-jets using MC particle-level flavor labeling. The B template is an aggregation of $bl$ and $bc$ events in which the leading track-jets are labeled as $b$ and the subleading track-jets are labeled as light-flavor or $c$ in particle-level flavor labeling.  This template also includes cases in which a $g\rightarrow b\bar{b}$ splitting was fully contained inside one small-radius track-jet and the second track-jet is due to a light quark or gluon. The rest of the events are merged into the L+C template.  Uncertainties in the templates, in particular resulting from the template merging scheme, are described in Section~\ref{sec:systs:exp}.

Examples of the flavor-fraction determination fits are shown in Figure~\ref{fig:fits} for one bin of $\Delta R(b,b)$.  Since the leading track-jet is required to be $b$-tagged, the distribution of $\subsdzero(j_1)$ is broader and shifted toward more positive values than the distribution of $\subsdzero(j_2)$.  In contrast, the subleading jet is most often a light-flavor jet for both the B and L+C categories and therefore $\subsdzero(j_2)$ is nearly symmetric about zero.  The BB template is similar in shape for the leading and subleading track-jets.  For this particular fit, the $\chi^2$ per degree of freedom improves from $72/22$ (pre-fit) to $13.5/22$ (post-fit).  As a result of the fit, the background fraction changes from 79.6\% to 82.8\%.  A comparison of the pre-fit and post-fit flavor fractions for all bins of $\Delta R$ as well as the other observables is presented in Figure~\ref{fig:dR-fitfrac}.  Except for $\Delta\mathrm{\theta}_\text{ppg,gbb}$ and the highest bin of $z(p_\text{T})$ the \PYTHIA simulation prediction for the shape of the flavor-fraction distribution is accurate for the BB fraction.  
In all cases the BB yield is slightly overestimated. The flavor fractions from B and L+C are inverted between \PYTHIA and data.

\begin{figure}[htbp]
  \centering
 \includegraphics[width=0.48\textwidth]{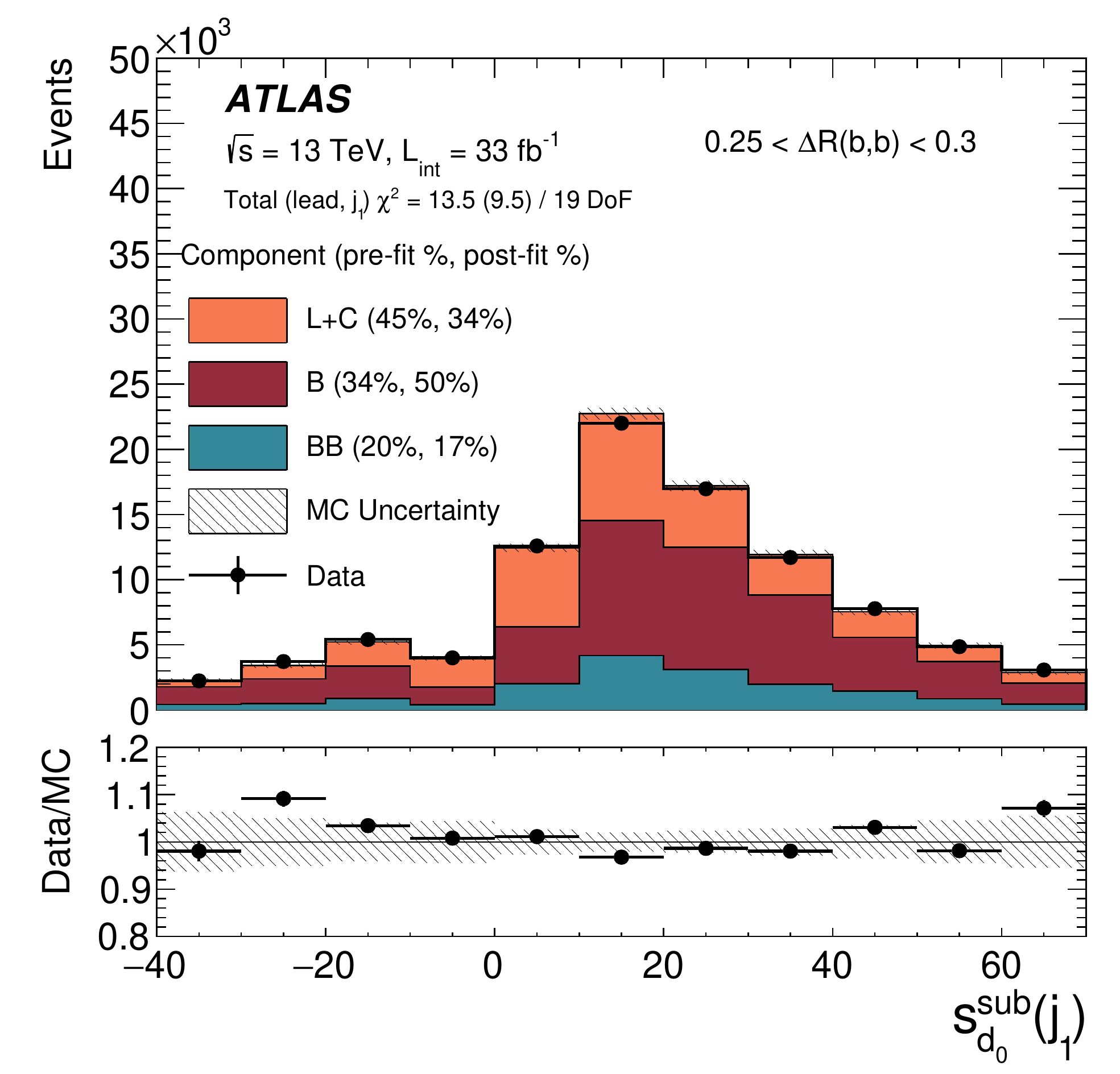}
 \includegraphics[width=0.48\textwidth]{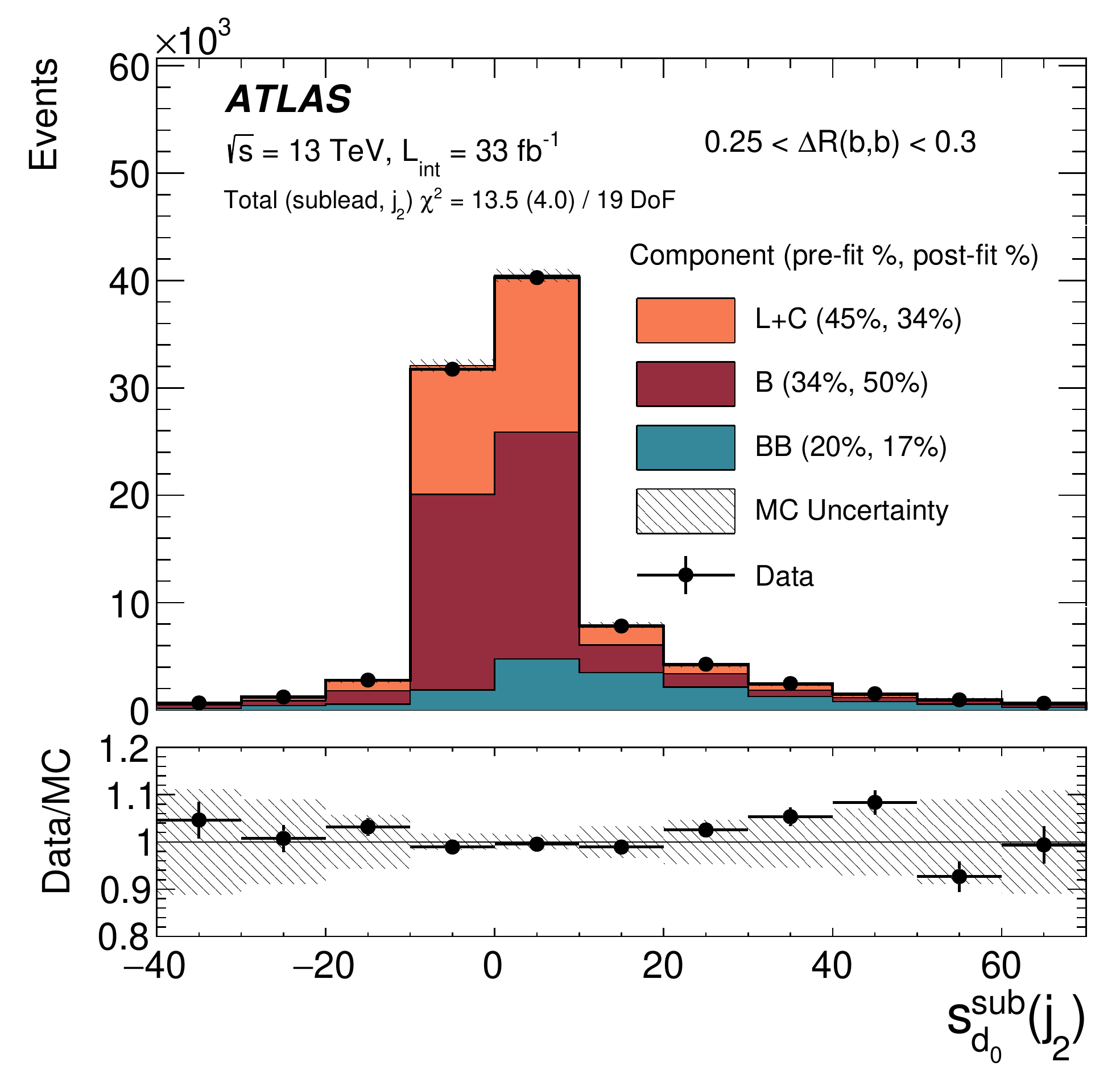}\\
\caption{The distribution of $\subsdzero$ in data and in simulation, post-fit, for the higher-$p_\text{T}$ track-jet (left) and for the lower-$p_\text{T}$ track-jet (right) in the bin $0.25<\Delta R(b,b)<0.3$. The three components are the signal double-$b$ (`BB'), the background single $b$ (`B'), and the background non-$b$ components (`L+C').  Percentages reported in the legend indicate the pre- and post-fit fraction of each component.  Only data and MC statistical uncertainties are shown.  The lower panel shows the ratio between data and the post-fit simulation.}
  \label{fig:fits}
\end{figure}

\begin{figure}[htbp]
  \centering
 \includegraphics[width=0.4\textwidth]{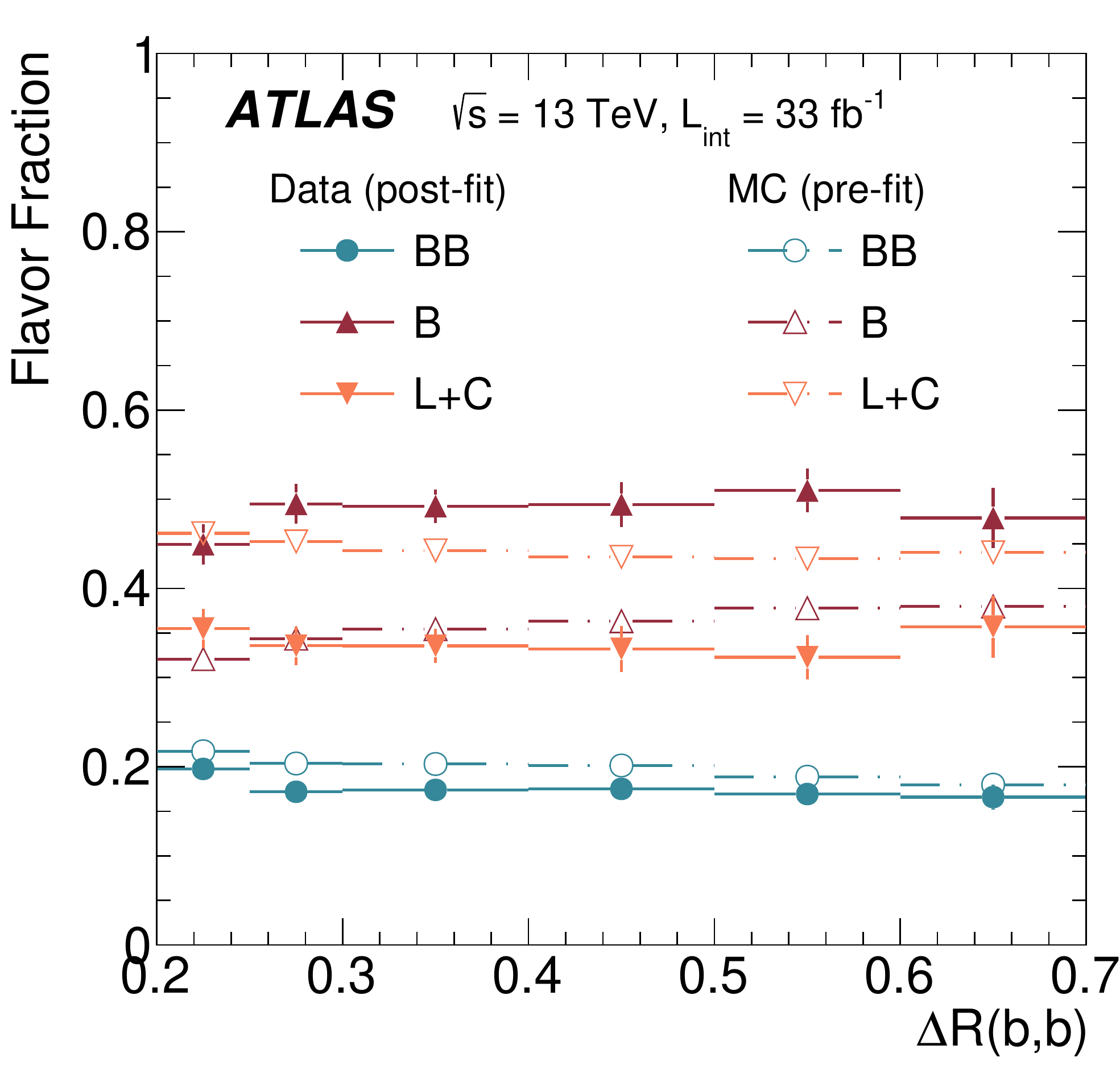}
 \includegraphics[width=0.4\textwidth]{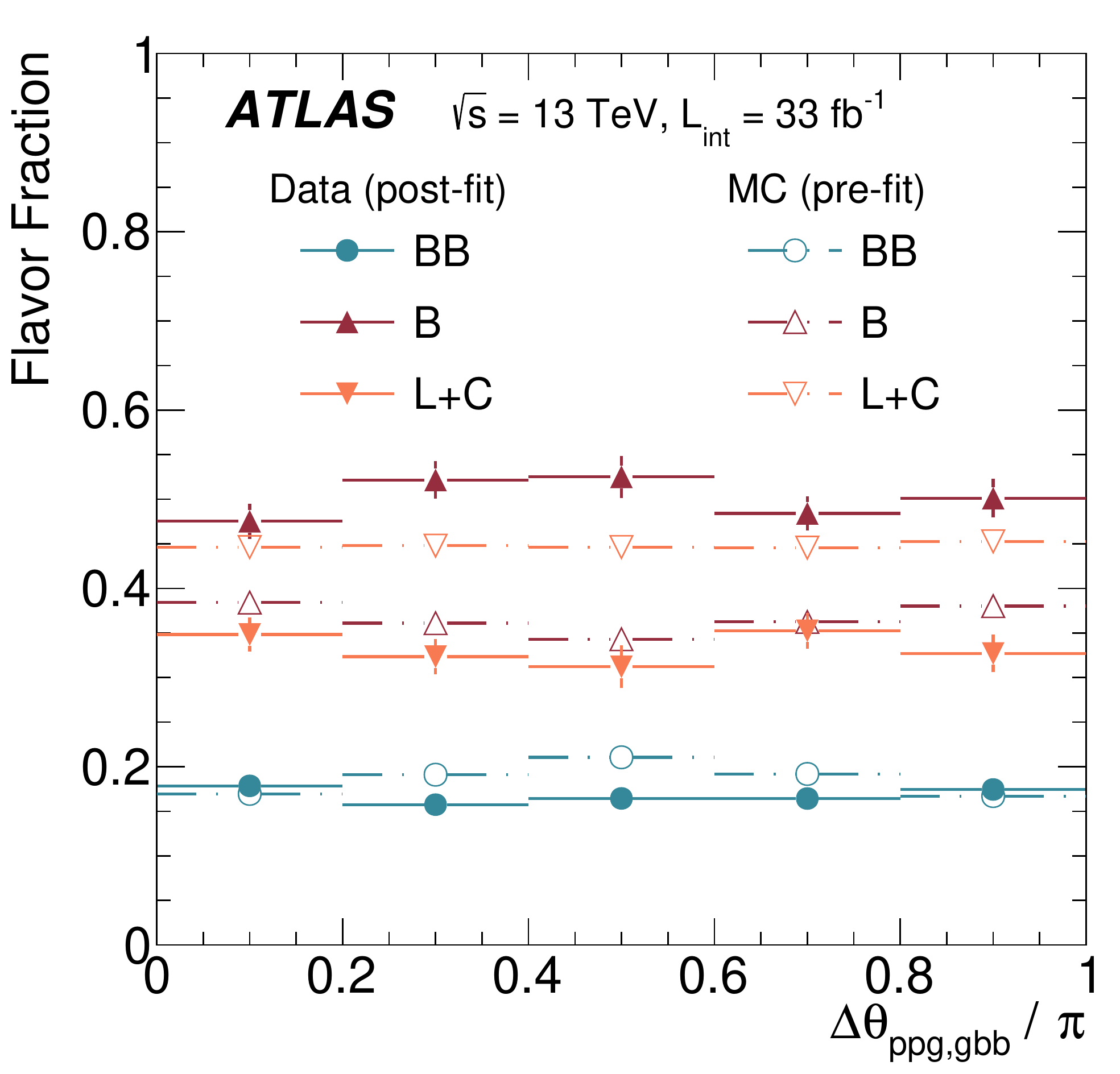}\\
 \includegraphics[width=0.4\textwidth]{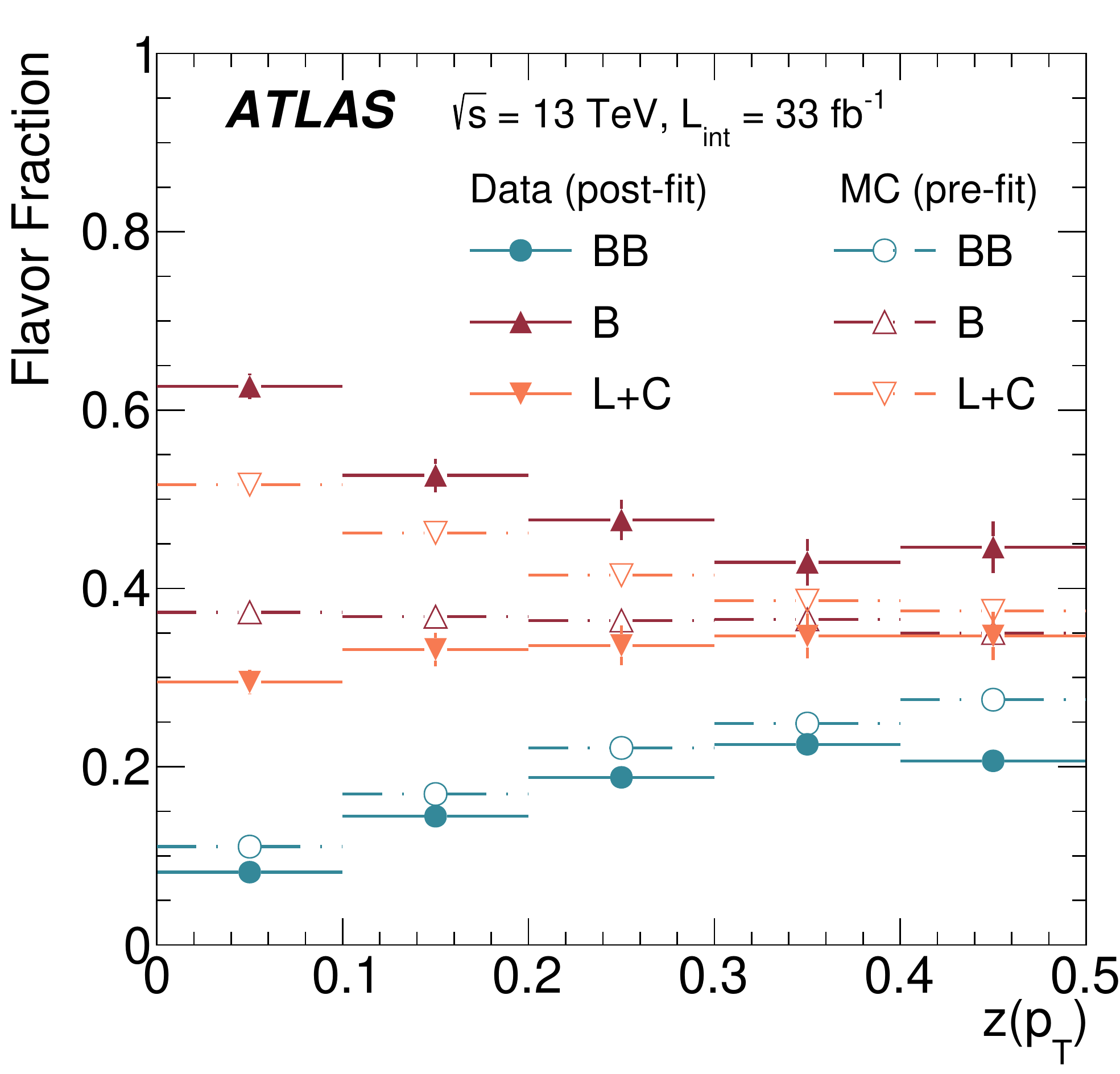}
 \includegraphics[width=0.4\textwidth]{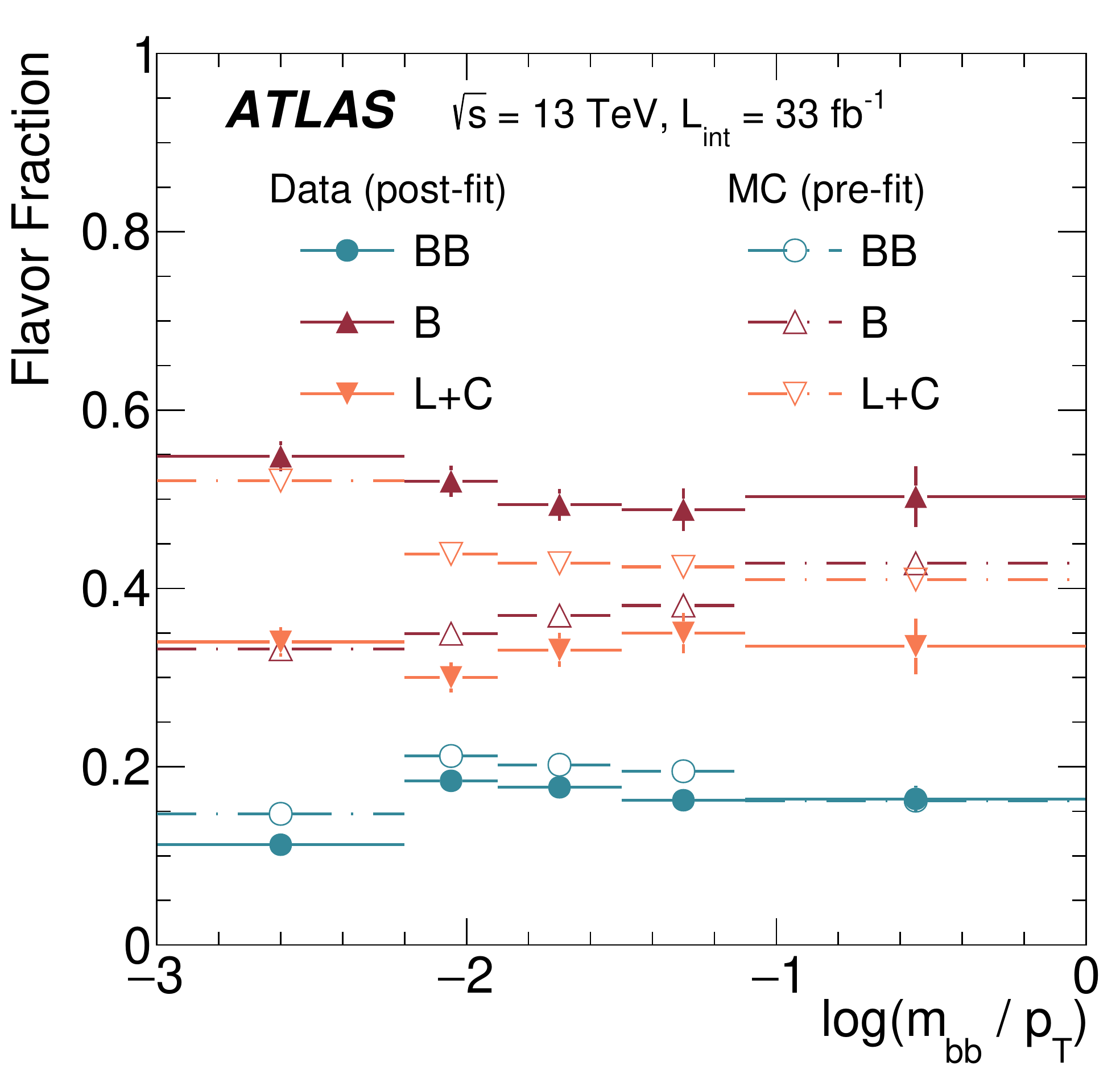}\\
\caption{The pre-fit (MC) and post-fit (data) flavor fractions for $\Delta R(b,b)$ (top left), $\Delta\mathrm{\theta}_\text{ppg,gbb}$ (top right), $z(p_\text{T})$ (bottom left), and $\log(m_{bb}/p_\text{T})$ (bottom right) are indicated with open and solid markers respectively. The error bars include only statistical uncertainties from the flavor-fraction fit.  The fit's systematic uncertainties are comparable in magnitude, but correlated across the bins.  The impact of both the flavor-fraction fit's statistical and systematic uncertainties on the final results is presented in Table~\ref{tab:syst}.}
  \label{fig:dR-fitfrac}
\end{figure}

\section{Unfolding}
\label{sec:unfold}

After subtracting the background from the detector-level distributions, as described in Section~\ref{sec:analysis}, the data are corrected for resolution and acceptance effects.  The fiducial volume of the measurement is described by the particle-level object and event selection in Section~\ref{sec:obj}.  First, the data are corrected for events that pass the detector-level selection but not the particle-level selection using the simulations introduced in Section~\ref{sec:data}.  Then, the iterative Bayes (IB) unfolding technique~\cite{DAgostini:1994zf} is used to correct for the detector resolution in events that pass both the detector-level and particle-level selections.  The IB method is applied with four iterations implemented in the RooUnfold framework~\cite{Adye:2011gm}.  After the application of the response matrix, a final correction is applied to account for events that pass the particle-level but not detector-level selection.  Uncertainties in the unfolding procedure are described in Section~\ref{sec:systs}.

\section{Uncertainties}
\label{sec:systs}

Systematic and statistical uncertainties were assessed for each aspect of the analysis, including the background subtraction, acceptance and efficiency correction factors, response matrix, and unfolding method.  For each uncertainty, a component of the analysis chain is varied and then the entire procedure including the background subtraction is repeated.  Table~\ref{tab:syst} provides a summary of the systematic uncertainties for each observable.  The jet energy scale, the unfolding, and the theoretical modeling  uncertainties dominate. Further details about each category are provided in Section~\ref{sec:systs:exp} for the uncertainties associated with each analysis object, in Section~\ref{sec:systs:background} for the background fit procedure uncertainty, and in Section~\ref{sec:systs:theory} for the unfolding method and theoretical modeling uncertainties.

\begin{table}[htbp]
\centering
\caption{Summary of systematic uncertainty sizes for each observable for the normalized differential cross sections.} 
\label{tab:syst}
\begin{tabular}{lrrrr}
\toprule
                       & $\Delta R(b,b)$ & $\Delta\mathrm{\theta}_\text{ppg,gbb} $           & $z(p_\text{T})$              & $\log(m_{bb}/p_\text{T})$                     \\ \midrule
Calorimeter jet energy               & 2--3\%    & 2--3\%              & 2--6\%             & 2--4\%                    \\
Flavor tagging            & <1\%     & <1\%               & <1\%              & <1\%                     \\
Tracking               & 1--2\%      & 1--2\%              & 2--4\%             & 1--2\%                    \\
Background fit & 1\%      & 1\%                & 1--2\%             & 2\%                      \\
Unfolding method             & 2--3\%    & 2\%                & 2--4\%             & 2--5\%                    \\
Theoretical modeling & 3--10\%      & 2--13\%              & 3--10\%               & 4--11\%                      \\ \midrule
Statistical & 1\%     & 1\%             & 2\%            & 1\%                    \\ 
Total & 3--10\%      & 3--10\%            & 3--14\%         &4--12\%                    \\  
\bottomrule
\end{tabular}
\end{table}

\subsection{Object reconstruction}
\label{sec:systs:exp}

Each object used in the analysis has an associated uncertainty.  These uncertainties affect the acceptance factors and the response matrix, as well as the background fit templates. 

\textbf{Calorimeter jets:} The energies of large-radius jets are shifted and smeared to account for uncertainties in both the bias and variance of the reconstructed energy.  Jet energy scale uncertainties are determined by comparing calorimeter-based and tracker-based jet energy measurements in inclusive dijet events, and range from 2\% to 6\%~\cite{LargeRJES}.

\textbf{Flavor tagging:} Data/MC corrections (`scale-factors') and uncertainties in the $b$-tagging efficiencies and $c$-jet misidentification rates are determined from $t\bar t$ events~\cite{btagtrkjet}.  Light-flavor misidentification rates are studied using dijet events. The $b$-jet uncertainties are 5--10\%, while the $c$-jet uncertainties are 20\% and light-flavor jet uncertainties are about 50\%. Uncertainties for track-jets with $p_\text{T}>300$~\GeV\ are extrapolated from low-$p_{\text{T}}$ jets as there are too few events with high-$p_{\text{T}}$ jets for a proper calibration. Extrapolation uncertainties are evaluated by varying quantities such as impact parameter resolutions and descriptions of the detector material. The extrapolated uncertainties range from 20\% to 100\% depending on the $p_{\text{T}}$ and the flavor of the jet.  As the flavor fractions are constrained in situ, there is a significant reduction from the prior flavor-tagging uncertainty values described above.  In particular, there is little sensitivity to the background scale factors and there is no sensitivity to inclusive flavor-tagging scale factors for $b$-jets.  Residual $\Delta R(b,b)$-dependent scale factor uncertainties that account for differences in the modeling of isolated versus non-isolated $b$-jets are derived from dedicated performance studies~\cite{ATLAS-CONF-2016-039,ATLAS-CONF-2016-002}.
  
\textbf{Tracking:} Systematic uncertainties are estimated for the track reconstruction efficiency, fake rate, and track parameter scales and resolutions.  The main source of inclusive tracking inefficiency is multiple scattering in the ID, so the uncertainty is set by the accuracy with which the ID material is simulated~\cite{PERF-2015-07}.  This leads to a 0.5\% uncertainty for $|\eta| < 0.1$, which grows to $2.7\%$ at the end of the ID acceptance.  An additional source of inefficiency arises inside the high-multiplicity environment in the cores of jets due to silicon pixel and microstrip cluster merging.  The uncertainty in the modeling of this density-induced inefficiency is about 0.8\%~\cite{PERF-2015-08,ATL-PHYS-PUB-2016-007}.  Fake tracks are due to combinations from multiple charged particles.  The track selection described in Section~\ref{sec:obj} reduces the contribution of fake tracks to much less than 1\% with a relative uncertainty that is about 30\%~\cite{ATL-PHYS-PUB-2017-016}.  The track parameters that are most relevant for this analysis are $p_\text{T}$ and $d_0$.  Weak modes in the ID alignment cause a bias in the track sagitta that is corrected for using a dedicated calibration and the calibration uncertainty is propagated through to the measurement~\cite{ATL-PHYS-PUB-2015-051}.  The modeling of the $d_0$ scale and resolution was studied in $Z\rightarrow \mu^+ \mu^-$ events at low $p_\text{T}$ and in dijet events at high $p_\text{T}$~\cite{ATL-PHYS-PUB-2015-051} and is used to assign an uncertainty to the modeling of these important track properties.

\subsection{Background fit}
\label{sec:systs:background}

In addition to the fit validations described in Section~\ref{sec:analysis}, several aspects of the fit are varied in order to assess the uncertainty in the extracted flavor fractions.

\textbf{Fit range:} The nominal flavor fraction fit is performed for \subsdzero$\in[-40,70]$. In order to assess the impact of this choice and the sensitivity to the \subsdzero~tails, the fit is repeated while excluding the left and right tails of the distributions, corresponding to \subsdzero $\in [-30, 70]$ and \subsdzero $\in [-40, 60]$.

\textbf{Template merging scheme:} Merging background components to form three aggregated templates fixes the relative fractions of the template sub-components. The sensitivity of the fitted flavor fractions to this choice is estimated by varying each merged background component up or down by a factor of two. 
The fit range has a bigger impact on the uncertainty than the merging variations of the flavor fractions.

\subsection{Unfolding method and theoretical modeling}
\label{sec:systs:theory}

An uncertainty resulting from the unfolding method described in Section~\ref{sec:unfold} is determined by unfolding the prediction from a different simulation with the nominal procedure.  The alternative simulation is constructed by reweighting the nominal particle-level spectrum so that the simulated detector-level spectrum, obtained by propagating the reweighted particle-level spectrum through the response matrix, agrees well with the data.  The modified detector-level distribution is unfolded with the nominal response matrix and the difference between this and the reweighted particle-level spectrum is an indication of the bias due to the unfolding method (in particular, the choice of prior)~\cite{Malaescu:2009dm}.

The unfolded result can depend on the modeling of jet fragmentation through the background fit, the prior, the response matrix, and the correction factors.  The $s_{d_0}$ distribution does not strongly depend on the properties of the jet radiation pattern, but an uncertainty is determined by taking the fitted background using templates from \SHERPA instead of \PYTHIA.  Variations in the prior are already accounted for in the data-driven non-closure uncertainty described above.  The rest of the contributions are evaluated by comparing the result using \PYTHIA with the result using the alternative \SHERPA sample described in Section~\ref{sec:analysis}.  This comparison is decomposed into components corresponding to varying only the response matrix or only the initial/final correction factors.  Varying only one component at a time is possible by reweighting a component of the \PYTHIA simulation to match the \SHERPA simulation and then evaluating the relative difference in the unfolded result.  All of the components are added in quadrature to determine the total uncertainty due to fragmentation modeling.  Each component is treated as uncorrelated because the uncertainty is based on only two fragmentation models and therefore a potential reduced uncertainty from exploiting potentially unphysical correlations between kinematic properties impacting acceptance and substructure attributes impacting the response is avoided.

\section{Results}
\label{sec:result}

The unfolded results along with multiple parton shower MC predictions are presented in Figure~\ref{fig:results}.  By construction, $0.2\leq\Delta R\leq 1.0$ and the peak around $0.3$ is due to the radius of the track-jets.  Furthermore, $0\leq z(p_\text{T})\leq 0.5$, with some distortions to the natural distribution at low values due to the $p_\text{T}$ threshold applied to the small-radius track-jets.

The \SHERPA predictions are generally more accurate than those from \PYTHIA, although there are significant differences between both generators and the data at low mass, low $z(p_\text{T})$ and for all $\Delta\mathrm{\theta}_\text{ppg,gbb}$.  The $\Delta\mathrm{\theta}_\text{ppg,gbb}$ distribution in data appears to be inverted with respect to the one from \PYTHIA (with a minimum instead of maximum at $\pi/2$) while \SHERPA predicts a relatively uniform distribution.  For comparison, the figure contains a version of \PYTHIA\footnote{Due to the available settings in \PYTHIA, different versions were used for these comparisons.  It was checked that there is no significant difference in $\Delta\mathrm{\theta}_\text{ppg,gbb}$ between \PYTHIA versions 8.186 and 8.230.} with the azimuthal asymmetries induced by gluon polarization turned off.\footnote{\texttt{TimeShower:phiPolAsym = off} and \texttt{TimeShower:phiPolAsymHard = off}.}  This sample appears to be closer to \SHERPA and also to the data.  In general, the properties of gluon polarization inside unpolarized hadrons are largely unconstrained by experimental data (see e.g.\ Ref.~\cite{Bacchetta:2018ivt} and references therein).  This and future measurements of $\Delta\mathrm{\theta}_\text{ppg,gbb}$ may provide a new way to extract $p_\text{T}$-dependent parton distributions in order to better understand proton structure and further improve the precision of various cross-section calculations~\cite{Boer:2011kf}.

In addition to studying gluon production properties, $g\rightarrow b\bar{b}$ provides a handle on gluon fragmentation.  Due to the large $b$-quark mass and in general the large $m_{bb}$ mass that is possible after splitting, there are many formally equivalent model choices in describing gluon fragmentation.  For example, the scale at which the strong coupling constant acts (renormalization scale) may be better described as scaling with $m_{bb}^2$ instead of the \PYTHIA default $p_{\text{T},bb}^2$.  To illustrate the sensitivity of $\Delta R(b,b)$, $z(p_\text{T})$ and  $\log(m/p_\text{T})$ to fragmentation settings in \PYTHIA, the plots of Figure~\ref{fig:results} show the final-state radiation variations of the A14 tune (indicated as an uncertainty band in the plot) as well as a different way to treat the $b$-quark mass in the QCD splitting kernels indicated by the $m_{bb}^2/4$ variation ($m_{bb}^2/4$ instead of $p_{\text{T},bb}^2$ for the renormalization scale).\footnote{\texttt{TimeShower:weightGluonToQuark = 5} and \texttt{TimeShower:scaleGluonToQuark = 0.25}. }  No variation describes all of the data, but some variations are worse than others.  For example, the Var2+ A14 variation, which increases the final-state shower $\alphas(M_Z)$ value to 0.139, moves the prediction further from the data in nearly all measurement bins. Related variations (not shown) such as using $m_{bb}^2$ instead of $p_{\text{T},bb}^2$ as the renormalization scale, adding additional phase-space factors, or suppressing high-mass $b\bar{b}$ pairs, are not significantly different from the nominal \PYTHIA\ setup.\footnote{Variations of \texttt{TimeShower:weightGluonToQuark} and \texttt{TimeShower:scaleGluonToQuark};  when \texttt{TimeShower:weightGluonToQuark = 1}, then the $g\rightarrow b\bar{b}$ kernel is weighted by an extra $\beta$ phase-space factor; when \texttt{TimeShower:weightGluonToQuark = 5}, this kernel is reweighted to $\alphas(m_{bb}^2)$ instead of $\alphas(p_\text{T}^2)$; when \texttt{TimeShower:weightGluonToQuark = 8}, there is an additional factor that suppresses the rate of high-mass $b\bar{b}$ pairs.}

\begin{figure}[htbp]
  \centering
 \includegraphics[width=0.49\textwidth]{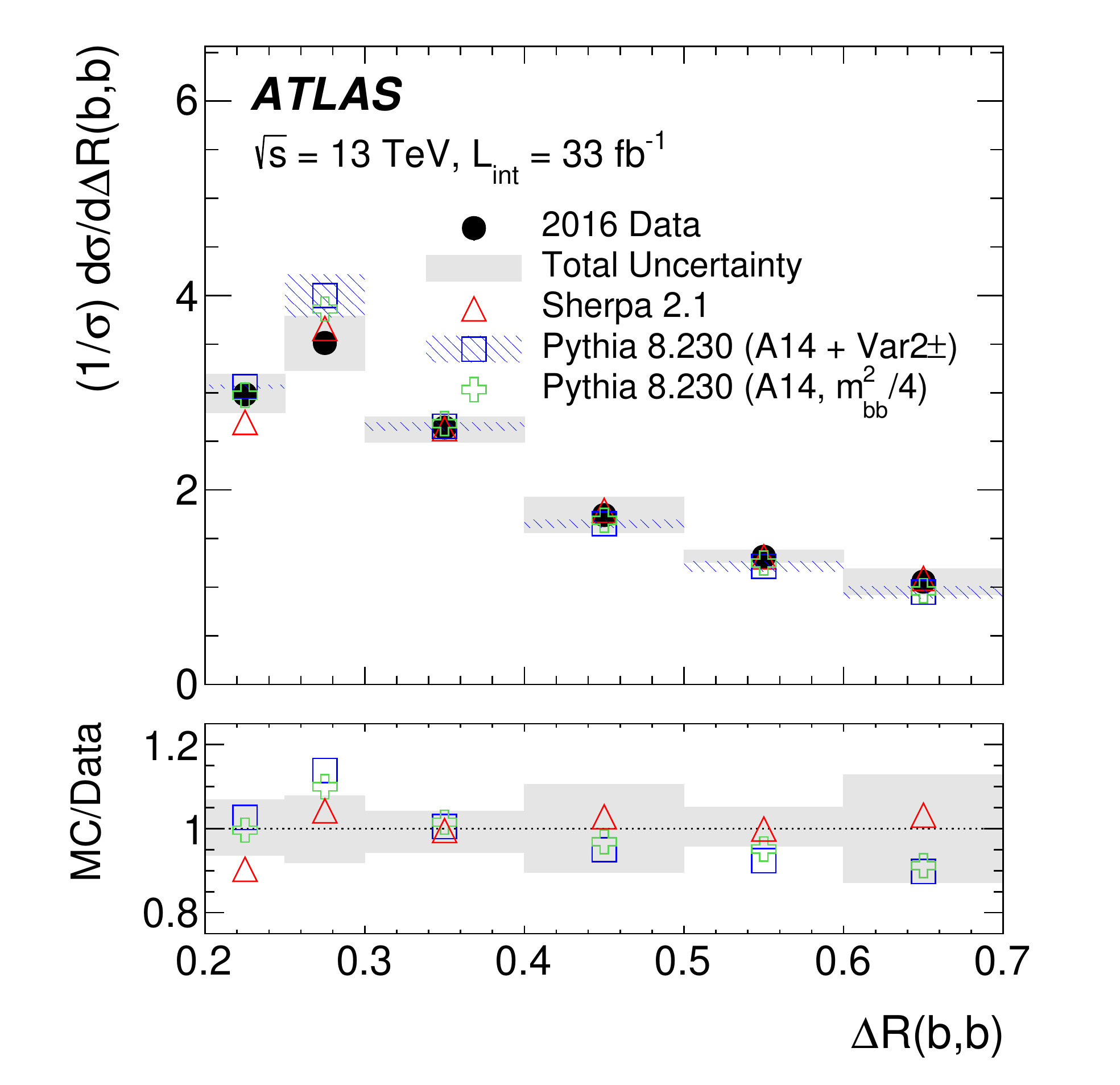} \includegraphics[width=0.49\textwidth]{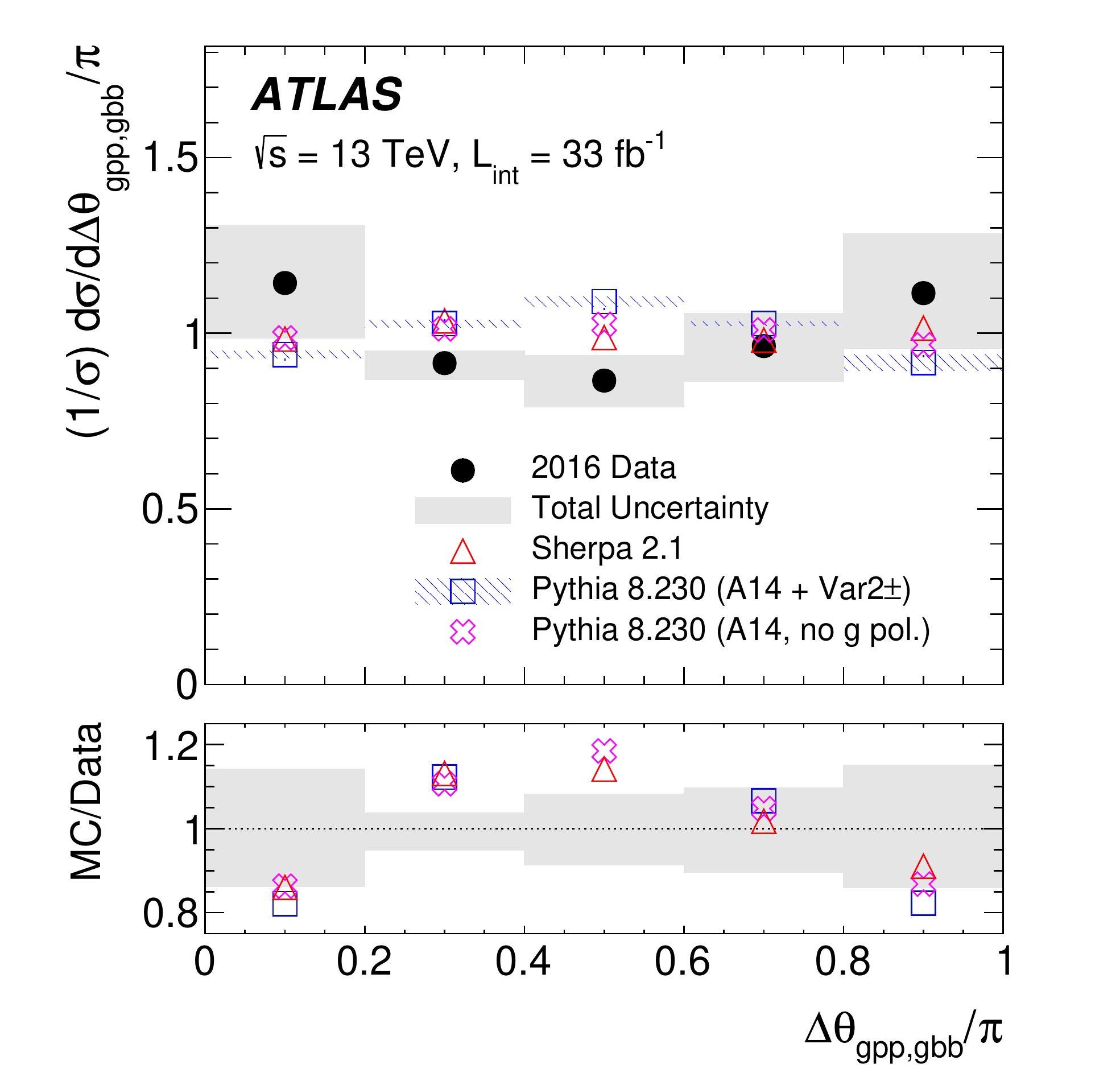}
 \includegraphics[width=0.49\textwidth]{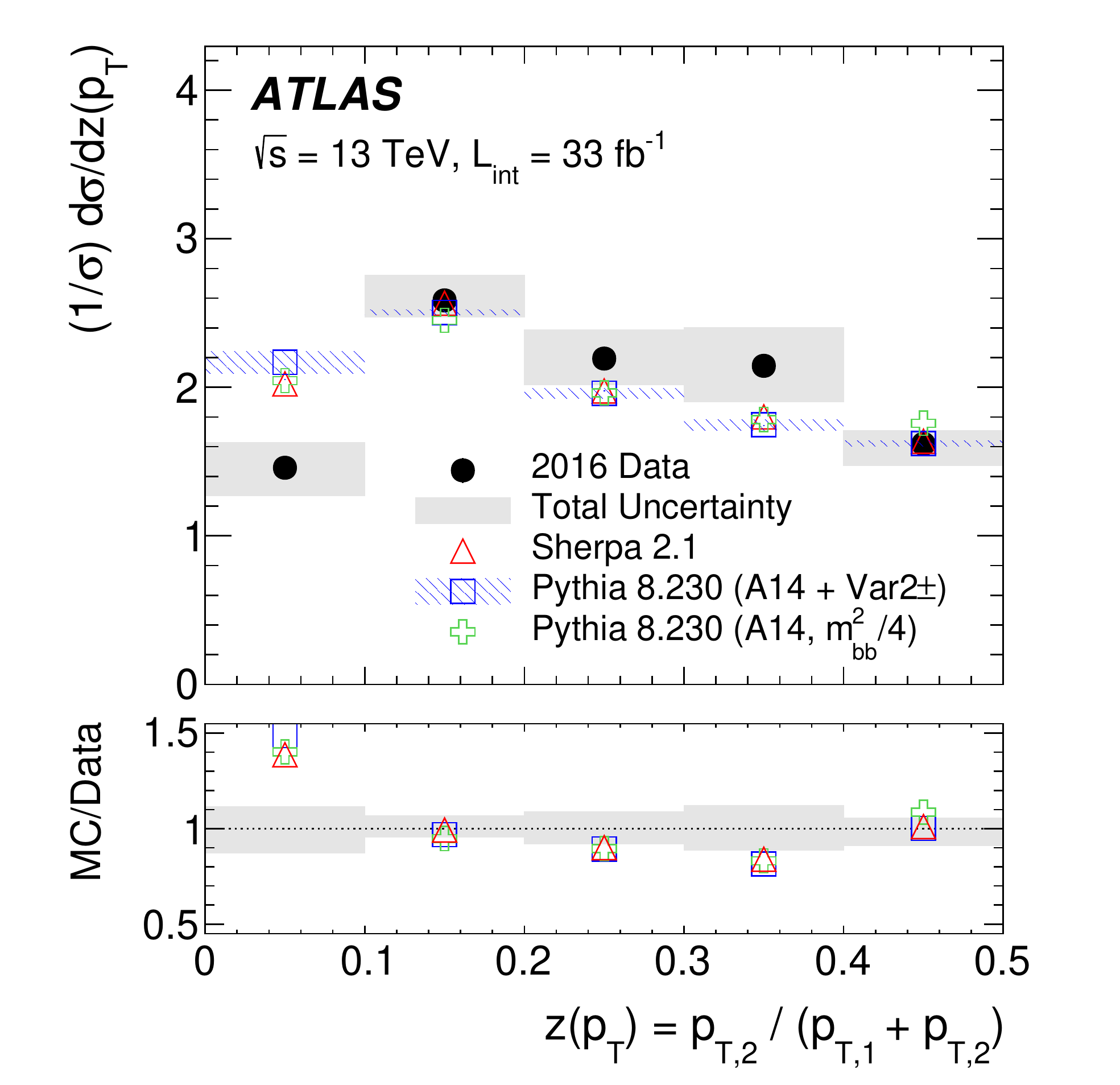} \includegraphics[width=0.49\textwidth]{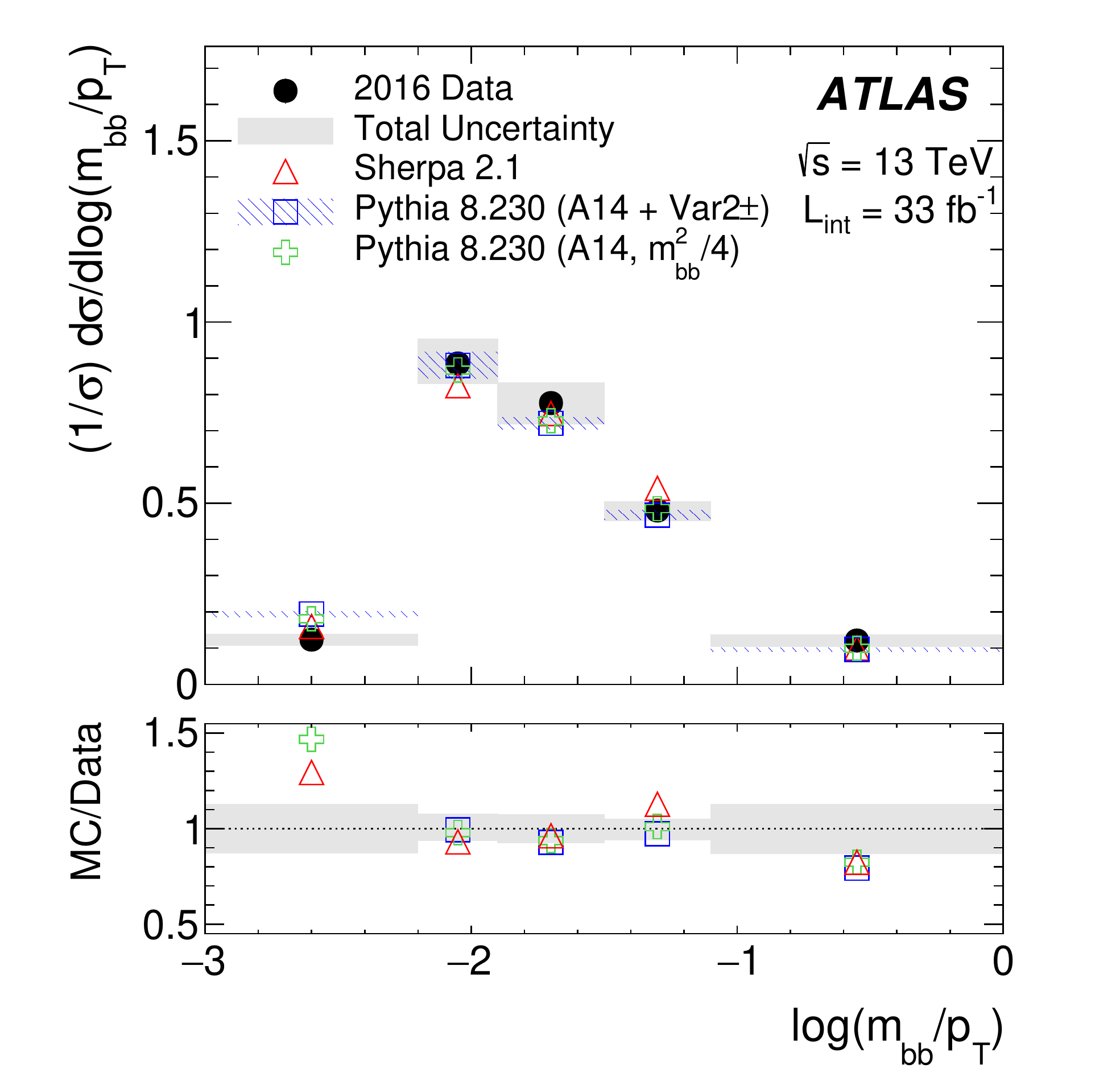}\\
\caption{The unfolded distribution of $\Delta R(b,b)$ (top left), $\Delta\mathrm{\theta}_\text{ppg,gbb}$ (top right), $z(p_\text{T})$ (bottom left), and $\log(m_{bb}/p_\text{T})$ (bottom right).  Error bands represent the sum in quadrature of statistical and systematic uncertainties (see Section~\ref{sec:systs}).  These data are compared with predictions from the \PYTHIA and \SHERPA MC simulations.  The bands for the \PYTHIA prediction represented by a square indicate the Var2$\pm$ variations (dominated by a $\pm 10\%$ variation in the final state shower $\alpha_s$).  The additional set of \PYTHIA markers use $m_{bb}^2/4$ for the renormalization scale.}
  \label{fig:results}
\end{figure}

\FloatBarrier

\section{Conclusion}
\label{sec:conclusion}

This paper presents a measurement of various properties of $g\rightarrow b\bar{b}$ at high $p_\text{T}$ and low $\Delta R(b,b)$ from 33~fb$^{-1}$ of $\sqrt{s}=13$~\TeV\ $pp$ collisions recorded by the ATLAS detector at the LHC.  A flavor-fraction fit is used to remove contributions from processes other than $g\rightarrow b\bar{b}$.  The fitted fractions significantly disagree with the pre-fit \PYTHIA predictions and suggest that further studies could improve the modeling of analyses sensitive to these fractions.  The measured properties are unfolded to correct for the detector acceptance and resolution for direct comparison with particle-level models.  Comparisons are made at particle level between the distributions and various models of jet formation. Simulations from the \SHERPA event generator generally provide a better model than \PYTHIA, especially for the $\Delta\mathrm{\theta}_\text{ppg,gbb}$ observable, which is sensitive to the modeling of the gluon polarization. The particle-level spectra are publicly available~\cite{hepdata} for further interpretation and can be used to validate QCD MC predictions and tune their models' free parameters.

\section*{Acknowledgments}


We thank CERN for the very successful operation of the LHC, as well as the
support staff from our institutions without whom ATLAS could not be
operated efficiently.

We acknowledge the support of ANPCyT, Argentina; YerPhI, Armenia; ARC, Australia; BMWFW and FWF, Austria; ANAS, Azerbaijan; SSTC, Belarus; CNPq and FAPESP, Brazil; NSERC, NRC and CFI, Canada; CERN; CONICYT, Chile; CAS, MOST and NSFC, China; COLCIENCIAS, Colombia; MSMT CR, MPO CR and VSC CR, Czech Republic; DNRF and DNSRC, Denmark; IN2P3-CNRS, CEA-DRF/IRFU, France; SRNSFG, Georgia; BMBF, HGF, and MPG, Germany; GSRT, Greece; RGC, Hong Kong SAR, China; ISF and Benoziyo Center, Israel; INFN, Italy; MEXT and JSPS, Japan; CNRST, Morocco; NWO, Netherlands; RCN, Norway; MNiSW and NCN, Poland; FCT, Portugal; MNE/IFA, Romania; MES of Russia and NRC KI, Russian Federation; JINR; MESTD, Serbia; MSSR, Slovakia; ARRS and MIZ\v{S}, Slovenia; DST/NRF, South Africa; MINECO, Spain; SRC and Wallenberg Foundation, Sweden; SERI, SNSF and Cantons of Bern and Geneva, Switzerland; MOST, Taiwan; TAEK, Turkey; STFC, United Kingdom; DOE and NSF, United States of America. In addition, individual groups and members have received support from BCKDF, CANARIE, CRC and Compute Canada, Canada; COST, ERC, ERDF, Horizon 2020, and Marie Sk{\l}odowska-Curie Actions, European Union; Investissements d' Avenir Labex and Idex, ANR, France; DFG and AvH Foundation, Germany; Herakleitos, Thales and Aristeia programmes co-financed by EU-ESF and the Greek NSRF, Greece; BSF-NSF and GIF, Israel; CERCA Programme Generalitat de Catalunya, Spain; The Royal Society and Leverhulme Trust, United Kingdom. 

The crucial computing support from all WLCG partners is acknowledged gratefully, in particular from CERN, the ATLAS Tier-1 facilities at TRIUMF (Canada), NDGF (Denmark, Norway, Sweden), CC-IN2P3 (France), KIT/GridKA (Germany), INFN-CNAF (Italy), NL-T1 (Netherlands), PIC (Spain), ASGC (Taiwan), RAL (UK) and BNL (USA), the Tier-2 facilities worldwide and large non-WLCG resource providers. Major contributors of computing resources are listed in Ref.~\cite{ATL-GEN-PUB-2016-002}.

\clearpage

\printbibliography

\clearpage
\input{atlas_authlist.tex}

\end{document}

%% file: atlas_authlist.tex
 
\begin{flushleft}
{\Large The ATLAS Collaboration}

\bigskip

M.~Aaboud$^\textrm{\scriptsize 34d}$,    
G.~Aad$^\textrm{\scriptsize 99}$,    
B.~Abbott$^\textrm{\scriptsize 125}$,    
D.C.~Abbott$^\textrm{\scriptsize 100}$,    
O.~Abdinov$^\textrm{\scriptsize 13,*}$,    
B.~Abeloos$^\textrm{\scriptsize 129}$,    
D.K.~Abhayasinghe$^\textrm{\scriptsize 91}$,    
S.H.~Abidi$^\textrm{\scriptsize 164}$,    
O.S.~AbouZeid$^\textrm{\scriptsize 39}$,    
N.L.~Abraham$^\textrm{\scriptsize 153}$
% ...
\bigskip
\\

% ...
$^{13}$Institute of Physics, Azerbaijan Academy of Sciences, Baku; Azerbaijan.\\
% ...
$^{34}$$^{(a)}$Facult\'e des Sciences Ain Chock, R\'eseau Universitaire de Physique des Hautes Energies - Universit\'e Hassan II, Casablanca;$^{(b)}$Centre National de l'Energie des Sciences Techniques Nucleaires (CNESTEN), Rabat;$^{(c)}$Facult\'e des Sciences Semlalia, Universit\'e Cadi Ayyad, LPHEA-Marrakech;$^{(d)}$Facult\'e des Sciences, Universit\'e Mohamed Premier and LPTPM, Oujda;$^{(e)}$Facult\'e des sciences, Universit\'e Mohammed V, Rabat; Morocco.\\
% ...
$^{39}$Niels Bohr Institute, University of Copenhagen, Copenhagen; Denmark.\\
% ...
$^{91}$Department of Physics, Royal Holloway University of London, Egham; United Kingdom.\\
% ...
$^{99}$CPPM, Aix-Marseille Universit\'e, CNRS/IN2P3, Marseille; France.\\
$^{100}$Department of Physics, University of Massachusetts, Amherst MA; United States of America.\\
% ...
$^{125}$Homer L. Dodge Department of Physics and Astronomy, University of Oklahoma, Norman OK; United States of America.\\
% ...
$^{129}$LAL, Universit\'e Paris-Sud, CNRS/IN2P3, Universit\'e Paris-Saclay, Orsay; France.\\
% ...
$^{153}$Department of Physics and Astronomy, University of Sussex, Brighton; United Kingdom.\\
% ...
$^{164}$Department of Physics, University of Toronto, Toronto ON; Canada.\\
% ...
$^{a}$ Also at Borough of Manhattan Community College, City University of New York, NY; United States of America.\\
$^{b}$ Also at California State University, East Bay; United States of America.\\
$^{c}$ Also at Centre for High Performance Computing, CSIR Campus, Rosebank, Cape Town; South Africa.\\
$^{d}$ Also at CERN, Geneva; Switzerland.\\
$^{e}$ Also at CPPM, Aix-Marseille Universit\'e, CNRS/IN2P3, Marseille; France.\\
% ...
$^{*}$ Deceased

\end{flushleft}
